\documentclass[twocolumn,showpacs,preprintnumbers,amsmath,amssymb]{revtex4}

\usepackage{graphicx}
\usepackage{dcolumn}
\usepackage{bm}
\usepackage{psfrag}

\def\h{\eta}
\def\th{\theta}

\begin{document}

\preprint{Brewster.tex}

\title{Geometric visualization of the Brewster angle 
from dielectric--magnetic interface}

\author{Ari Sihvola}
 \email{ari.sihvola@tkk.fi}
\affiliation{%
Helsinki University of Technology, Electromagnetics Laboratory\\
P.O. Box 3000, Espoo, FIN-02015, Finland
}%

 \homepage{http://www.tkk.fi/~asihvola}

\date{\today}

\begin{abstract}
A geometric visualization is presented for the Brewster angle for a
plane wave reflecting from an interface. The surface is assumed
isotropic but it is allowed to display both dielectric and magnetic
susceptibility, and hence the Brewster (polarizing) angle can attain
any value between 0 and 90 degrees, and can exist for both parallel
and perpendicular polarizations. The geometric construction (a
tetrahedron) is spanned by the basic material parameters of the
surface. The Brewster angle appears in one of the faces of the
tetrahedron.
\end{abstract}


\maketitle

\section{Introduction}

Very much physics is sometimes contained in simple and basic results
of optics and electromagnetics. In this paper I shall focus on the
character of electromagnetic waves reflected from a planar surface. As
is well known, many everyday light phenomena that we can observe with
plain eyes \cite{Minnaert} can be justified and explained with basic
wave theory which is is being taught to freshmen in physics and
engineering schools. As examples in optics we could mention the glare
on road surfaces on a sunny day which can be reduced by use of
Polaroid sun glasses, or the way the images reflected from a water
surface differ from those that are direcly observed.

The polarization state of light changes in refrection and refraction
processes. Since our eyes are not capable of sensing polarization, and
natural light very often is rather unpolarized, the subtleties of the
outdoor images, as they appear to us, may only be present in very
indirect ways. But one especially interesting phenomenon in this
respect is the possibility of light to become fully polarized in
reflection. This happens when light impinges on a surface in a certain
direction, from the Brewster angle. In the following, let us
concentrate on the dependence of Brewster angle on the fundamental
material parameters. In particular, the emphasis shall be on the way
how the Brewster angle can be visualized in a geometrical way which
contains pedagogical and physical insight.

In the following, the materials to be analyzed are assumed isotropic
and lossless. However, in one respect the analysis is more general
than that encountered in basic textbooks in optics which often
restrict the treatment to non-magnetic media: here also magnetic
permeability is taken as a material parameter that can vary. Presently
in many engineering applications, composite materials research, and
nanotechnology, great interest is in the magnetic properties of
matter, which gives motivation to allow magnetic contrasts in the
studies of canonical problems.

Hence, if both electric and magnetic responses are present, the
material from which the wave reflects is characterized by two
parameters, the relative permittivity and permeability $\epsilon$ and
$\mu$. These are assumed in the present paper to be real and positive
\footnote{It is perhaps important to emphasize here the explicit assumption
of positiveness of the material parameters. In recent years very much
research has been and is still being focused on materials with
negative permittivity and permeability values, so-called
negative-phase-velocity media, left-handed media, or metamaterials
\cite{Veselago,Pendry}. Large research programs have been launched in
the U.S.\ and in Europe which target on design and exploitation of
metamaterials; see, for example, {\tt
http://www.darpa.mil/dso/thrust/matdev/metamat.htm} and {\tt
http://www.metamorphose-eu.org}}. But to ease the analysis, instead of
using these parameters, it appears more convenient to apply the
refractive index $n$ and relative impedance $\h$ of the material:
\begin{equation}
n = \sqrt{\epsilon\mu}, \quad \h = \sqrt{\mu/\epsilon}
\end{equation}
Obviously the inverse relations are $\epsilon=n/\h$ and $\mu=n\h$.

The following sections give the reflection coefficients from such a
material and a way to visualize them.

\section{Reflection coefficients}

\begin{figure}[h]
\psfragscanon 
\psfrag{t1}[][]{{$\th_1$}}
\psfrag{t2}[][]{{$\th_2$}}
\psfrag{nh}[][]{{$n,\quad \h$}}
\centerline{\includegraphics[width=8cm]{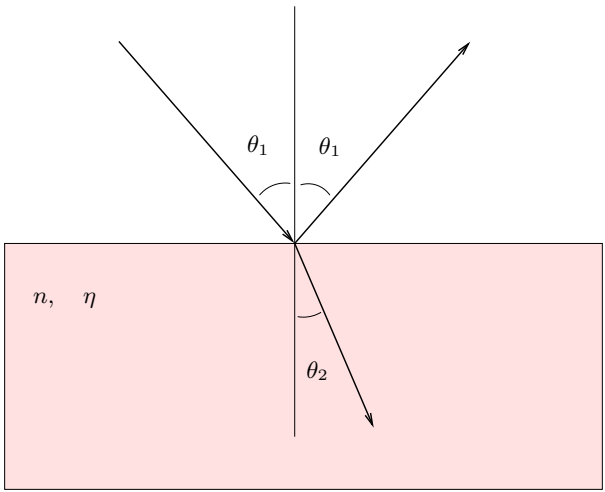}}
  \caption{Plane wave hitting a boundary between free space and a
  dielectric--magnetic material with refractive index $n$ and
  impedance $\h$.}
\label{fig:iim1}
\end{figure}

The geometry of the problem to be analyzed is very simple and shown in
Figure~\ref{fig:iim1}. An incident electromagnetic wave is impinging
from free space and faces a planar interface. On the other side of the
boundary, there is a homogeneous half space of dielectric--magnetic
medium with refractive index and impedance parameters $n$ and
$\h$. After the collision with the boundary, part of the energy is
refracted and penetrates into the medium, and the remaining part
reflects away form the interface.

In general, the wave changes its polarization state in
reflection. Only for two eigenpolarizations of the incident wave do
the reflected and refracted waves remain with the same polarization as
the incoming wave. These two are parallel (P) and perpendicular (S)
polarizations, meaning that the linearly polarized electric field
vector is in the plane of incidence (P) or perpendicular to it
(S). The plane of incidence is spanned by the incident wave direction
and the normal of the interface (the plane of paper in
Figure~\ref{fig:iim1}).

The reflection coefficients for the two polarizations can be written
in many equivalent forms \cite{Jackson,Born_Wolf}; the following
electric field Fresnel coefficients are quite symmetric:
\begin{eqnarray}
R_{\rm P} & = & \frac{\h \cos\th_2 - \cos\th_1}{\h \cos\th_2 +
\cos\th_1} \label{1} \\ 
R_{\rm S} & = & \frac{\h \cos\th_1 - \cos\th_2}{\h
\cos\th_1 + \cos\th_2} \label{2}
\end{eqnarray} 
In using these formulas, the value for the refraction angle 
$\th_2$ is needed. It is determined by the Snell's law
\begin{equation}\label{3}
\sin\th_1 = n \sin\th_2
\end{equation}

These expressions give the reflected electric field vector for unit
incident field. The magnitudes of the reflection coefficients are
always between zero and unity. Note, however, that the reflection
coefficients can attain complex values even in the case of real values
for $n$ and $\h$; this happens for total internal reflection with the
associated Goos--H\"anchen phenomenon.

\begin{figure}[h!]
\psfragscanon 
\psfrag{n}[][]{{$n$}}
\psfrag{h}[][]{{$\h$}}
\psfrag{1}[][]{{$1$}}
\psfrag{P}[][]{{{\tt P}}}
\psfrag{S}[][]{{\tt S}}
\centerline{\includegraphics[width=8cm]{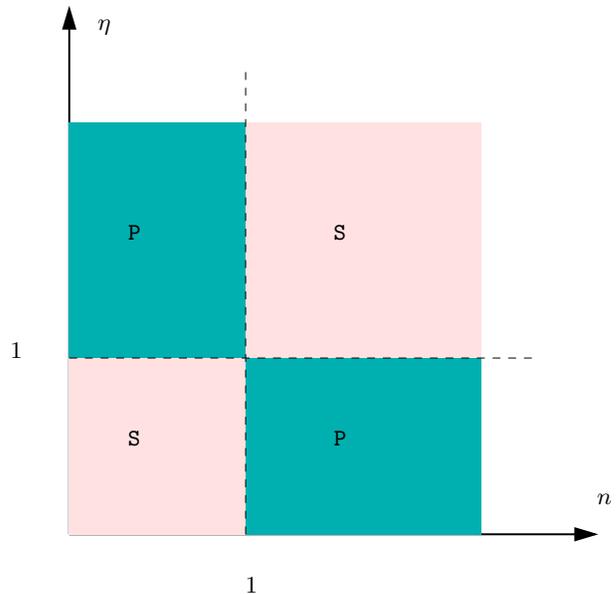}}
  \caption{Regions of the ($n$-$\h$)-plane where the Brewster angle
  can be observed for parallel and perpendicular polarizations.}
\label{fig:plane1}
\end{figure}

Of course, very interesting is the case when the reflection
vanishes. It is easy to solve from (\ref{1})--(\ref{3}) the incidence
angle for which the reflection coefficient is zero. This is called the
Brewster angle, and it is for the parallel polarization
\begin{equation}\label{BrP}
\th_{\rm Br,P} = \arcsin \left( n \sqrt{\frac{1-\h^2}{n^2-\h^2}} \right)
\end{equation}
For the perpendicular polarization the Brewster angle can be written as
\begin{equation}\label{BrS}
\th_{\rm Br,S} = \arcsin \left( n \sqrt{\frac{\h^2-1}{n^2\h^2-1}} \right)
\end{equation}
Note that only for one polarization there exists a Brewster angle; the
requirements are (see Figure~\ref{fig:plane1})
\begin{itemize}
\item Parallel polarization: $n>1$ and $\h<1$,  or  $n<1$ and $\h>1$
\item Perpendicular polarization: $n>1$ and $\h>1$,  or  $n<1$ and $\h<1$
\end{itemize}

Note that the expression (\ref{BrP}) is a generalization from the
familiar Brewster-angle relation $\tan\th_{\rm Br,P}=n$ which is valid
for non-magnetic media $(\h=1/n)$, and naturally only exists for the
parallel polarization. When magnetic response is allowed, the relation
for the polarizing angle has one more degree of freedom. It can be
written, of course, also in forms other than (\ref{BrP})--(\ref{BrS}),
see, for example \cite{Futterman}.

\begin{figure}[h]
\centerline{\includegraphics[width=9cm]{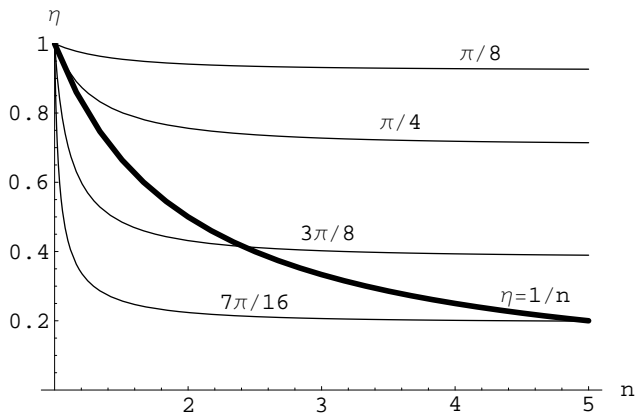}}
  \caption{Equi-Brewster-angle curves in the ($n$-$\h$) -plane for
  parallel polarization. Four curves are shown. The thick curve
  $\h=1/n$ divides the plane into a upper ``paramagnetic part'' where
  $\mu>1$, and the lower ``diamagnetic part'' where $\mu<1$.}
\label{fig:mmm30}
\end{figure}

An interesting observation is that the Brewster angle can attain any
values between zero and 90$^\circ$, as can be seen from
Figure~\ref{fig:mmm30} in case of parallel polarization. Note that
for ordinary dielectric materials where $n=1/\h$ the Brewster angle
$\th_{\rm Br}=\arctan(n)$ is larger than 45$^\circ$. For the parallel
polarization, the impedance as function of the refractive index and
the Brewster angle is
\begin{equation}
\h = \frac{n \cos\th_{\rm Br}}{\sqrt{n^2 - \sin^2\th_{\rm Br}}}
\end{equation}

The simple law for the non-magnetic Brewster angle $\tan\th_1=n$,
combined with the Snell's law $\sin\th_1=n\sin\th_2$ yields
$\cos\th_1=\sin\th_2$. This means that the incidence and refracted
angles are complementary angles $(\th_1+\th_2=90^\circ)$. Therefore
(see Figure~\ref{fig:iim1}) the direction of the reflected wave is
orthogonal to the refracted wave. In such a geometric constellation
the dipoles induced in the medium by the refracted ray, which have a
radiation null along their axis direction, do not cause reradiation
into the direction of the reflected ray. Hence physical intuition
agrees with the result of Brewster angle formula \cite{Sastry,DeSmet},
although the interpretation has been also criticized
\cite{Nitzan,Merzbacher}.

But let us return to the more general case of the properties of the
wave that reflects from a dielectric--magnetic interface.

\section{Geometric interpretation}

The square roots of differences of squares in the relations
(\ref{BrP}) and (\ref{BrS}) for the two Brewster angles remind of the
Pythagorean theorem. And indeed, after some time of trigonometric play
with these relations, beautiful geometric interpretations can be
discovered from right triangles that are built from the three basic
measures $n$, $\h$, and $n\h$. Further, an arrangement of these
triangles in three dimensions reveals structures with which the
Brewster angles can be grasped in a very visual sense.

\begin{figure*}[h!]
\psfragscanon 
\psfrag{t}[][]{{$\th_{\rm Br,P}$}}
\psfrag{1}[][]{{$n$}}
\psfrag{2}[][]{{$\h$}}
\psfrag{3}[][]{{$n\h$}}
\psfrag{4}[][]{{$n\sqrt{1-\h^2}$}}
\psfrag{5}[][]{{$\h\sqrt{n^2-1}$}}
\psfrag{6}[][]{{$\sqrt{n^2-\h^2}$}}
\psfrag{X}[][]{{$\begin{array}{ll}{\bf Parallel\ polarization}\\ 
(n>1,\;\; \h<1) \end{array}$}}
\psfrag{X2}[][]{{$\begin{array}{ll}{\bf Parallel\ polarization}\\ 
(n<1,\;\; \h>1) \end{array}$}}
\centerline{\includegraphics[width=13cm]{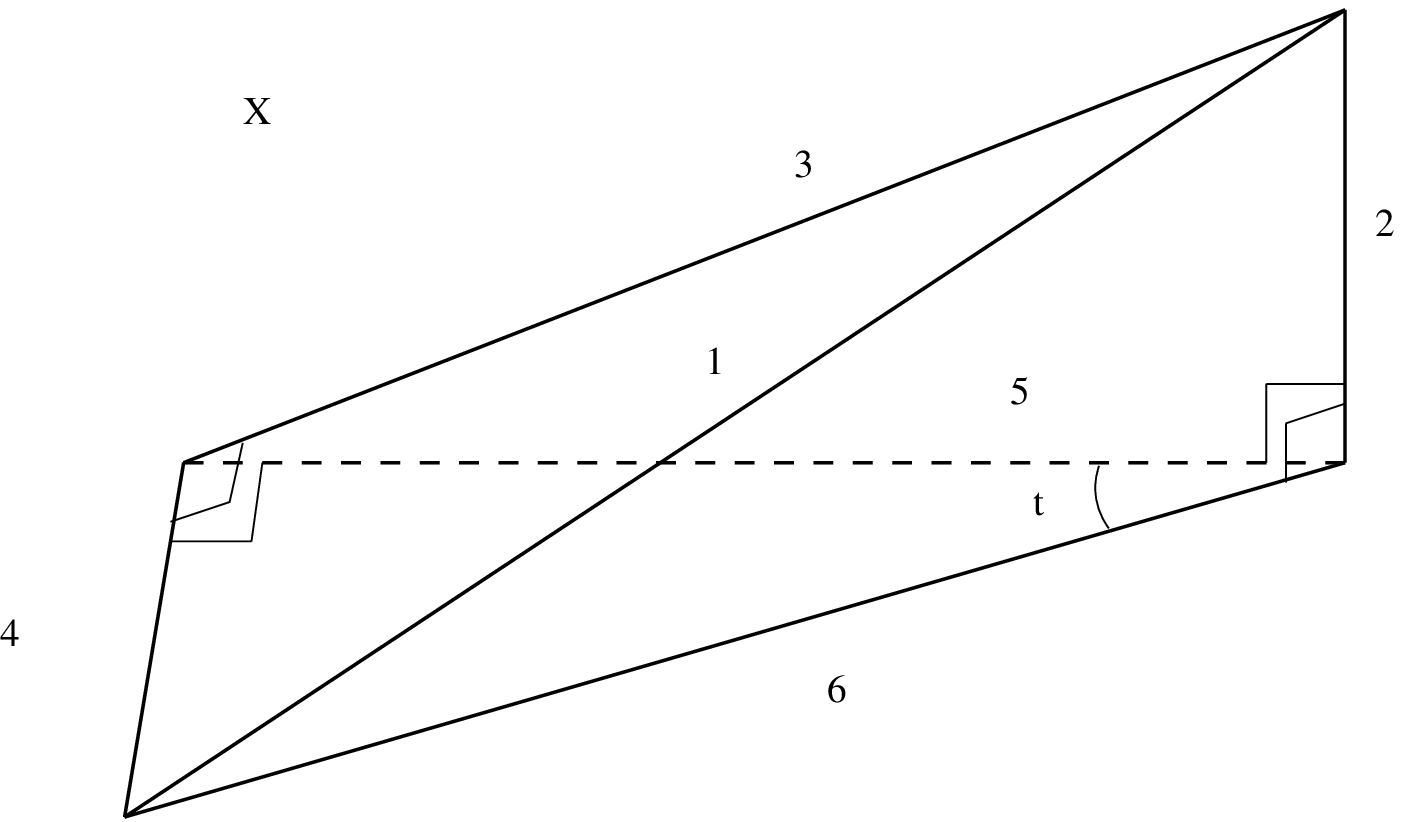}}
\vspace{10mm}
\psfrag{1}[][]{{$\h$}}
\psfrag{2}[][]{{$n$}}
\psfrag{3}[][]{{$\h n$}}
\psfrag{4}[][]{{$\h\sqrt{1-n^2}$}}
\psfrag{5}[][]{{$n\sqrt{\h^2-1}$}}
\psfrag{6}[][]{{$\sqrt{\h^2-n^2}$}}
\centerline{\includegraphics[width=13cm]{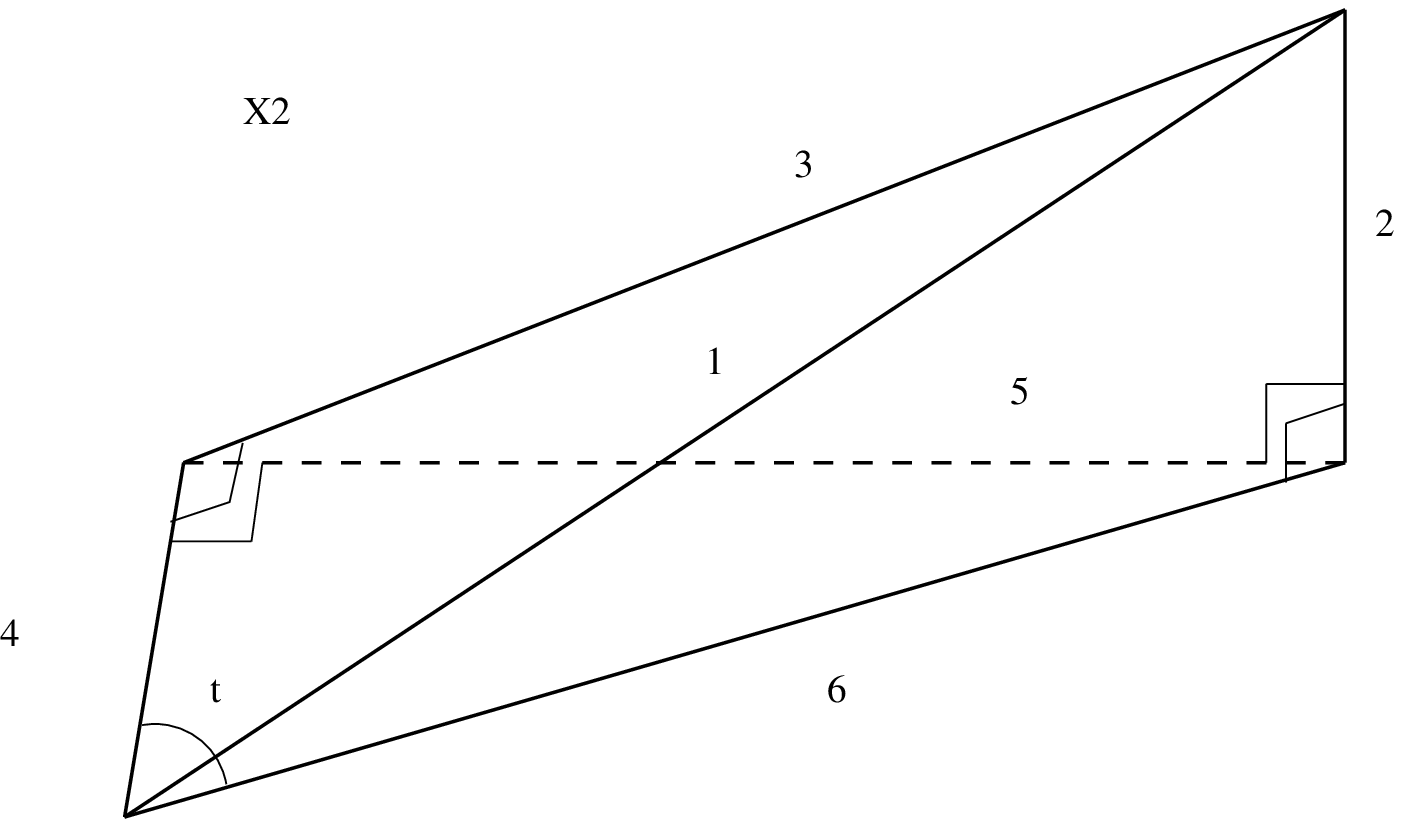}}
  \caption{A geometrical view of the Brewster angle determined by the
  primary material constants $n$ and $\h$. Parallel polarization,
  $n>1,\h<1$ (upper figure); $n<1,\h>1$ (lower figure). Note the four
  right-triangular faces of the tetrahedra.}
\label{fig:tetra}
\end{figure*}

This geometric construction is illustrated in Figure~\ref{fig:tetra}
for the relations expressing the Brewster angle for parallel
polarization. From the magnitudes of $n$ and $\h$, a tetrahedron is
uniquely determined. The faces of this geometrical object are four
right triangles. The Brewster angle can be read from the bottom of the
tetdahedron.

Figure~\ref{fig:tetra2} shows the same for the perpendicular
polarization.

\begin{figure*}[h]
\psfragscanon 
\psfrag{t}[][]{{$\th_{\rm Br,S}$}}
\psfrag{1}[][]{{$n\h$}}
\psfrag{2}[][]{{$1$}}
\psfrag{3}[][]{{$n$}}
\psfrag{4}[][]{{$n\sqrt{\h^2-1}$}}
\psfrag{5}[][]{{$\sqrt{n^2-1}$}}
\psfrag{6}[][]{{$\sqrt{n^2\h^2-1}$}}
\psfrag{X}[][]{{$\begin{array}{ll}{\bf Perpendicular\ polarization}\\ 
(n>1,\;\; \h>1) \end{array}$}}
\psfrag{X2}[][]{{$\begin{array}{ll}{\bf Perpendicular\ polarization}\\ 
(n<1,\;\; \h<1) \end{array}$}}
\centerline{\includegraphics[width=13cm]{tetraBr.eps}}
\vspace{10mm}
\psfrag{1}[][]{{$1$}}
\psfrag{2}[][]{{$n\h$}}
\psfrag{3}[][]{{$n$}}
\psfrag{4}[][]{{$\sqrt{1-n^2}$}}
\psfrag{5}[][]{{$n\sqrt{1-\h^2}$}}
\psfrag{6}[][]{{$\sqrt{1-\h^2n^2}$}}
\centerline{\includegraphics[width=13cm]{tetraBr2.eps}}
  \caption{The same as in Figure~\ref{fig:tetra}, for the perpendicular
  polarization. Upper figure: $n>1,\h>1$; lower figure: $n<1,\h<1$.}
\label{fig:tetra2}
\end{figure*}

\section{Conclusion}

Sir David Brewster performed his studies on the character of reflected
light during the second decade of the 19th century. Therefore the
concept of polarizing angle is nearly as old as the understanding of
the transverse nature of light. The fascinating manner how the
material properties affect the appearance of the Brewster angle is
very interesting still today, both from experimental application point
of view and also pedagogically when we are learning physics, optics,
and electromagnetism. Hopefully the present article can give a helpful
contribution to a modern understanding of the Brewster angle.

\bibliography{apssamp}

\begin{thebibliography}{10}
\expandafter\ifx\csname natexlab\endcsname\relax\def\natexlab#1{#1}\fi
\expandafter\ifx\csname bibnamefont\endcsname\relax
  \def\bibnamefont#1{#1}\fi
\expandafter\ifx\csname bibfnamefont\endcsname\relax
  \def\bibfnamefont#1{#1}\fi
\expandafter\ifx\csname citenamefont\endcsname\relax
  \def\citenamefont#1{#1}\fi
\expandafter\ifx\csname url\endcsname\relax
  \def\url#1{\texttt{#1}}\fi
\expandafter\ifx\csname urlprefix\endcsname\relax\def\urlprefix{URL }\fi
\providecommand{\bibinfo}[2]{#2}
\providecommand{\eprint}[2][]{\url{#2}}

\bibitem[{\citenamefont{Minnaert}(1993)}]{Minnaert}
\bibinfo{author}{\bibfnamefont{M.~G.~J.} \bibnamefont{Minnaert}},
  \emph{\bibinfo{title}{Light and Color in the Outdoors}}
  (\bibinfo{publisher}{Springer-Verlag}, \bibinfo{year}{1993}).

\bibitem[{\citenamefont{Jackson}(1999)}]{Jackson}
\bibinfo{author}{\bibfnamefont{J.~D.} \bibnamefont{Jackson}},
  \emph{\bibinfo{title}{Classical Electrodynamics}} (\bibinfo{publisher}{John
  Wiley \& Sons}, \bibinfo{year}{1999}), \bibinfo{note}{third edition}.

\bibitem[{\citenamefont{Born and Wolf}(1980)}]{Born_Wolf}
\bibinfo{author}{\bibfnamefont{M.}~\bibnamefont{Born}} \bibnamefont{and}
  \bibinfo{author}{\bibfnamefont{E.}~\bibnamefont{Wolf}},
  \emph{\bibinfo{title}{Principles of Optics}} (\bibinfo{publisher}{Pergamon
  Press}, \bibinfo{year}{1980}), \bibinfo{note}{sixth edition}.

\bibitem[{\citenamefont{Futterman}(1995)}]{Futterman}
\bibinfo{author}{\bibfnamefont{J.}~\bibnamefont{Futterman}},
  \bibinfo{journal}{American Journal of Physics} \textbf{\bibinfo{volume}{63}},
  \bibinfo{pages}{471} (\bibinfo{year}{1995}).

\bibitem[{\citenamefont{Sastry and Chakrabarty}(1984)}]{Sastry}
\bibinfo{author}{\bibfnamefont{G.~P.} \bibnamefont{Sastry}} \bibnamefont{and}
  \bibinfo{author}{\bibfnamefont{S.}~\bibnamefont{Chakrabarty}},
  \bibinfo{journal}{American Journal of Physics} \textbf{\bibinfo{volume}{52}},
  \bibinfo{pages}{177} (\bibinfo{year}{1984}).

\bibitem[{\citenamefont{DeSmet}(1994)}]{DeSmet}
\bibinfo{author}{\bibfnamefont{D.~J.} \bibnamefont{DeSmet}},
  \bibinfo{journal}{American Journal of Physics} \textbf{\bibinfo{volume}{62}},
  \bibinfo{pages}{246} (\bibinfo{year}{1994}).

\bibitem[{\citenamefont{Nitzan and Bodenheimer}(1984)}]{Nitzan}
\bibinfo{author}{\bibfnamefont{M.}~\bibnamefont{Nitzan}} \bibnamefont{and}
  \bibinfo{author}{\bibfnamefont{J.~S.} \bibnamefont{Bodenheimer}},
  \bibinfo{journal}{American Journal of Physics} \textbf{\bibinfo{volume}{52}},
  \bibinfo{pages}{660} (\bibinfo{year}{1984}).

\bibitem[{\citenamefont{Merzbacher}(1985)}]{Merzbacher}
\bibinfo{author}{\bibfnamefont{E.}~\bibnamefont{Merzbacher}},
  \bibinfo{journal}{American Journal of Physics} \textbf{\bibinfo{volume}{53}},
  \bibinfo{pages}{916} (\bibinfo{year}{1985}).

\bibitem[{\citenamefont{Veselago}(1968)}]{Veselago}
\bibinfo{author}{\bibfnamefont{V.~G.} \bibnamefont{Veselago}},
  \bibinfo{journal}{Soviet Physics Uspekhi} \textbf{\bibinfo{volume}{10}},
  \bibinfo{pages}{509} (\bibinfo{year}{1968}).

\bibitem[{\citenamefont{Pendry}(2000)}]{Pendry}
\bibinfo{author}{\bibfnamefont{J.~P.} \bibnamefont{Pendry}},
  \bibinfo{journal}{Physical Review Letters} \textbf{\bibinfo{volume}{85}},
  \bibinfo{pages}{3966} (\bibinfo{year}{2000}).

\end{thebibliography}


\begin{thebibliography}{3}
\expandafter\ifx\csname natexlab\endcsname\relax\def\natexlab#1{#1}\fi
\expandafter\ifx\csname bibnamefont\endcsname\relax
  \def\bibnamefont#1{#1}\fi
\expandafter\ifx\csname bibfnamefont\endcsname\relax
  \def\bibfnamefont#1{#1}\fi
\expandafter\ifx\csname citenamefont\endcsname\relax
  \def\citenamefont#1{#1}\fi
\expandafter\ifx\csname url\endcsname\relax
  \def\url#1{\texttt{#1}}\fi
\expandafter\ifx\csname urlprefix\endcsname\relax\def\urlprefix{URL }\fi
\providecommand{\bibinfo}[2]{#2}
\providecommand{\eprint}[2][]{\url{#2}}

\bibitem[{\citenamefont{Feynman}(1954)}]{feyn54}
\bibinfo{author}{\bibfnamefont{R.~P.} \bibnamefont{Feynman}},
  \bibinfo{journal}{Phys.\ Rev.} \textbf{\bibinfo{volume}{94}},
  \bibinfo{pages}{262} (\bibinfo{year}{1954}).

\bibitem[{\citenamefont{Witten}()}]{witten2001}
\bibinfo{author}{\bibfnamefont{E.}~\bibnamefont{Witten}},
  \eprint{hep-th/0106109}.

\bibitem[{\citenamefont{Einstein et~al.}(1935)\citenamefont{Einstein, Podolsky,
  and Rosen}}]{epr}
\bibinfo{author}{\bibfnamefont{A.}~\bibnamefont{Einstein}},
  \bibinfo{author}{\bibfnamefont{B.}~\bibnamefont{Podolsky}}, \bibnamefont{and}
  \bibinfo{author}{\bibfnamefont{N.}~\bibnamefont{Rosen}},
  \bibinfo{journal}{Phys.\ Rev.} \textbf{\bibinfo{volume}{47}},
  \bibinfo{pages}{777} (\bibinfo{year}{1935}).

\end{thebibliography}

\end{document}